\documentclass[conference]{IEEEtran}
\IEEEoverridecommandlockouts
\usepackage{amsmath,amssymb,amsfonts}
\usepackage{algorithmic}
\usepackage{textcomp}
\usepackage{xcolor}
\def\BibTeX{{\rm B\kern-.05em{\sc i\kern-.025em b}\kern-.08em
    T\kern-.1667em\lower.7ex\hbox{E}\kern-.125emX}}
\PassOptionsToPackage{bookmarks={false}}{hyperref}      

\usepackage{color,soul}         

\usepackage{cite}                                    
\usepackage[pdftex]{graphicx,hyperref,color}      
\begin{document}

\title{A Combinational Multi-Kernel Decoder for Polar Codes}
\author{ \IEEEauthorblockN{{Hossein~Rezaei$^{1*}$,~Nandana~Rajatheva$^1$~,~Matti~Latva-aho$^1$~}}
\IEEEauthorblockA{$^1$Centre for Wireless Communications,~University of Oulu, Finland\\
E-mail: \{hossein.rezaei, nandana.rajatheva, matti.latva-aho\}@oulu.fi}}
\maketitle
\begin{abstract}Polar codes have been selected as the channel coding scheme for control channel in the fifth generation (5G) communication system thanks to their capacity achieving characteristics. However, the traditional polar codes support only codes constructed by binary ($2\times 2$) kernel which limits the code lengths to powers of $2$. Multi-kernel polar codes are proposed to achieve flexible block length. In this paper, the first combinational decoder for multi-kernel polar codes based on successive cancellation algorithm is proposed. The proposed decoder can decode pure-binary and binary-ternary ($3\times 3$) mixed polar codes. The architecture is rate-flexible with the capability of online rate assignment and supports any kernel sequences.  The FPGA implementation results reveal that for a code of length $N=48$, the coded throughput of $812.1$ Mbps can be achieved. 
\end{abstract}
\section{Introduction}
\IEEEPARstart{P}{olar} codes have been subjected to growing attention due to their capability to achieve symmetric channel capacity of binary-input discrete memoryless channels at infinite code length \cite{Arikan, Rezaei2022, 9621127}. During the last decade, researchers have extensively improved polar codes in terms of error-correction performance for finite-length codes, decoding latency under successive cancellation (SC) algorithm, complexity and power. This effort paved the way for polar codes to be adapted in the 3GPP fifth generation new radio (5G-NR) wireless communication standard \cite{3GPP}.

However, the majority of current research works have concentrated on polar codes constructed by binary kernels ($2\times2$ polarization matrix) also known as Arikan's kernel \cite{Arikan}. The lengths of polar codes are therefore restricted to powers of $2$. The 5G framework demands various code lengths and code rates. The rate-matching schemes \cite{Han2022}, \cite{bioglio2017} are proposed to address this limitation. However, a priori performance and optimality evaluation of these methods are hard. Multi-kernel (MK) polar codes \cite{Xia2020, gabry2017, bioglio} offer flexible code lengths with the same computational complexity as Arikan's polar codes by employing kernels with variable dimensions. It is shown in \cite{gabry2017} that MK polar codes outperform similar codes constructed by puncturing and shortening methods in terms of error-correction performance.  

Several architectures have been proposed for decoding Arikan's polar codes. 
Recently, an architecture for implementing MK polar codes constructed from binary and ternary ($3\times 3$) kernels have been proposed in \cite{Coppolino} by adapting the architecture of Arikan's codes in \cite{Sarkis} to MK codes. The work in \cite{Rezaei2022MK} proposed a MK architecture to reduce the latency of the MK polar codes. However, the coded throughput is not promising. Also, two mentioned MK decoders need different memory interfaces for binary and ternary stages which adds to the complexity of the memory system. 

In this paper, we propose an architecture based on the SC algorithm targeting high-throughput MK polar codes with low power consumption. The recursive and feed-forward structure of the SC algorithm makes use of pure combinational logic possible to implement the decoder's functions. The operation frequency of combinational decoders is lower than that of the sequential counterparts. However, they are capable of decoding an entire codeword in only one clock cycle resulting in considerable reduction in dynamic power with reference to sequential decoders. The proposed architecture features an online rate assignment mechanism for a given block length and it can decode pure-binary and binary-ternary mixed kernels. Since pure-ternary codes generate odd block lengths, they are out of our interest. An FPGA implementation is conducted to validate the architecture and the results are compared to the state-of-the-art MK decoders.

The remainder of this paper is organized as follows. A background on polar codes is presented in section \ref{sec_back}. Section \ref{sec_arch} details the code construction method, the proposed decoder's architecture and complexity analysis. The implementation results and comparison to previous works are summarized in section \ref{sec_res}. Finally, section \ref{sec_conc} concludes this work.
\section{Polar codes}
\label{sec_back}
A polar code of length $N$ carrying $K$ bits of information is denoted by $\mathcal{PC}(N,K)$ and the code rate is computed as $\mathcal{R} = \frac{K}{N}$. The set of $K$ information bits are called information set ($\mathcal{I}$) and the set of remaining $N-K$ bits are called frozen set ($\mathcal{F}$) which are set to $0$. Arikan proposed channel polarization phenomenon \cite{Arikan} to transform the physical channel $W$ into $N$ individual virtual channels $W_{i}^{N}$ ($1 \leq i \leq N$) with relative increased or decreased reliabilities. The reliability of each channel approaches either $0$ (completely unreliable) or $1$ (completely reliable) as the code lengths approaches infinity. The individual reliable channels can be designated by Bhattacharya parameters \cite{Arikan}.

A binary polar code can be constructed through a linear transformation expressed as $x = uG$. Here $x$ is the encoded stream, $u$ is a N-bit input vector to the encoder constructed by message and frozen data placement into reliable and unreliable positions, respectively. Finally, $G = T_2^{\otimes n}$ is the binary generator matrix constructed by the n-th Kronecker product of Arikan's kernel $T_2=[\begin{smallmatrix} 1&1\\ 1&0 \end{smallmatrix}]$. Obviously, $G$ is defined in a recursive way where a binary polar code of length $N$ is generated by concatenating two codes of length $N/2$. 
\subsection{MK Polar Codes}
Utilizing $T_2$ as the generator matrix bounds the block lengths of polar codes to powers of 2. However, using the LDPC WiMAx codelengths \cite{Shin2012} as guideline reveals that the code lengths constructed by non-binary kernels are needed. Most of the desired code lengths can be obtained using only one or few non-binary kernels. 
In order to construct a block code as $N = n_0\times n_1\times...\times n_s$ with $n_i$s being not necessarily individual prime numbers, a series of Kronecker products between different kernels can form the generator matrix as $G \triangleq T_{n_0}\otimes T_{n_1}\otimes...\otimes  T_{n_s}$, where $T_{n_i}$s are squared matrices.

Each distinctive prime number can be considered as a kernel, however the least complex and most practical kernels are binary and ternary (defined as $T_3=[\begin{smallmatrix} 1&1&1 \\ 1&0&1\\ 0&1&1  \end{smallmatrix}]$ \cite{gabry2017}) kernels. It is shown in \cite{benammar2017} that $T_3$ offers polarization optimality, although it has a lower polarization exponent than $T_2$. In this paper, we investigate codes constructed by pure-binary and any combination of binary and ternary kernels. The pure-ternary kernels are not of our interest since they generate an odd block length. 

The block length and the generator matrix for the polar codes presented in this paper can be formulated as $N=2^n.3^m$ and $G =  \otimes_{i=0}^{m+n} T_{k_i}$ where $n, m \in \mathbb{N}$. Defining the number of terms in the generator matrix as $M=m+n$ with $M \in [1,10]$, the increased length flexibility offered by MK polar codes is outlined in Table \ref{tab:codelength}. The bold values indicate codes with possible use cases in $5G$ polar codes.
\begin{table}
\centering
\caption{List of block lengths attainable by MK codes using $T_2$, $T_3$.}
\begin{tabular}{|c|cccccccccc|}
\hline
M &1&2&3&4&5&6&7&8&9&10 \\\hline
&2&4&8&16&\textbf{32}&\textbf{64}&\textbf{128}&\textbf{256}&\textbf{512}&\textbf{1024} \\
&3&6&12&24&\textbf{48}&\textbf{96}&\textbf{192}&\textbf{384}&\textbf{768}&1536\\
&&9&18&\textbf{36}&\textbf{72}&\textbf{144}&\textbf{288}&\textbf{576}&1152&2304\\
&&&27&\textbf{54}&\textbf{108}&\textbf{216}&\textbf{432}&\textbf{864}&1728&3456\\
&&&&\textbf{81}&\textbf{162}&\textbf{324}&\textbf{648}&1296&2592&5184\\
N&&&&&\textbf{243}&\textbf{486}&\textbf{972}&1944&3888&7776\\
&&&&&&\textbf{729}&1458&2916&5832&11664\\
&&&&&&&2187&4374&8748&17496\\
&&&&&&&&6561&13122&26244\\
&&&&&&&&&19682&39366\\
&&&&&&&&&&59048\\
\hline
\end{tabular}
\label{tab:codelength}
\end{table}

\begin{figure}
    \centering
    \includegraphics[width=1\columnwidth]{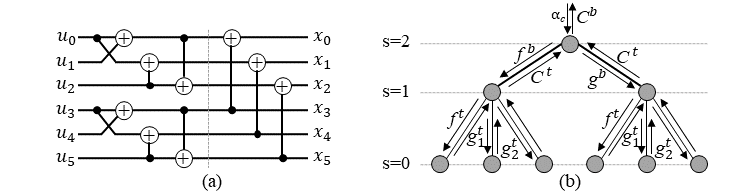}
    \caption{a) Encoder graph and b) decoder tree of a MK polar code of size $N=6$ with $G = T_2 \otimes T_3$.}
    \label{fig:EncDec}
\end{figure}

Lets take a simple MK polar code $N=6$ as an example. Two different kernel orders of $G=T_2\otimes T_3$ and $G=T_3\otimes T_2$ which generate unique generator matrices can be considered for this code. Since the Kronecker product is not commutative, different kernel orders generate distinctive polar codes with different error-correction performances. Fig. \ref{fig:EncDec} (a) illustrates the polarizing construction when $N=6$ and $G=T_2\otimes T_3$.
\subsection{MK Successive-Cancellation Decoding}
Arikan \cite{Arikan} proposed the SC algorithm to decode polar codes. MK polar codes can be decoded using the same algorithm. Fig. \ref{fig:EncDec} (b) depicts the decoder tree corresponding to the tanner graph of Fig. \ref{fig:EncDec} (a). The channel's soft information ($\alpha_c$) called log-likelihood ratios (LLRs) enter the tree from the root. To estimate a codeword, the soft information needs to propagate to the tree and visit all the leaves in a sequential way with the condition of visiting the left node first. Three functions are needed to traverse the tree. For a given binary node $\nu$, $\alpha_{v_l}$ is the function required to travel to the left branch which can be estimated as 
\begin{equation}
    \alpha^b_{v_l}[i]=sgn(\alpha_v[i] . \alpha_v[i + 2^{(\lambda-1)}])min(|\alpha_v[i]| , |\alpha_v[i + 2^{(\lambda-1)}]|) ~ 
    \label{eq:sc_l}
\end{equation}
where $i \in [0 : 2^{(\lambda-1)}-1]$. The node $\nu$ can compute the LLR vector to be sent to the right branch when the hard decisions ($\beta^b _{v_l}$) from the left branch are available. 
\begin{equation}
    \alpha^b_{v_r}[i]= (1-2\beta^b_{v_l}[i])\alpha_v[2i] + \alpha_v[2i+1]~ \textrm{for} ~i \in [0:2^{(\lambda-1)}-1].
    \label{eq:sc_r}
\end{equation}
where $\alpha^b_{v_r}$ is the LLR of the right branch. The codeword at node $\nu$ called $\beta^{\nu b}$ can be generated by combining at node $\nu$ when the hard decision bits of the right branch are available.
\begin{equation}
\begin{aligned}
\relax[\beta^{\nu b}_i, \beta^{\nu b}_{i + 2^{(\lambda-1)}}] ={} & [\beta^{\nu bl}_i\oplus\beta^{\nu br}_i,\beta^{\nu br}_i].
\label{eq:betak2}
\end{aligned}
\end{equation}
The hard decisions on a leaf node can be estimated as 
\begin{equation}
\begin{aligned}
\beta_v =     \begin{cases}
h(\alpha _v), & \text{if}\ v \in \mathcal{I} , \\
0, & \text{if}\ v \in \mathcal{F} 
\end{cases},\
h(x) =     \begin{cases}
0,   & \text{if}\ x \geq 0, \\
1,   & \text{otherwise.}
\end{cases}
\label{HD}
\end{aligned}
\end{equation}
We define (\ref{eq:sc_l}), (\ref{eq:sc_r}) and (\ref{eq:betak2}) as $f^b$, $g^b$ and $C^b$, respectively, as shown in Fig. \ref{fig:EncDec} (b).
The message passing criterion for a ternary node needs defining four functions. For a given node $\nu$ located at level $\lambda$ in a pure-ternary polar code, the decoding functions for traveling to the left, middle and right branches are shown by $\alpha^t_{v_l}$, $\alpha^t_{v_c}$ and $\alpha^t_{v_r}$, respectively. For $i \in ~[0, 3^{\lambda{\text -}1}{\text -}1]$ the $\alpha^t_{v_l}$ is calculated as 
\begin{equation}
\begin{aligned}
\alpha^t_{v_l}[i] ={} & sgn(\alpha_v[i] . \alpha_v[i + 2^{(\lambda-1)}]. \alpha_v[i + 2^{\lambda}]) \\
    & min(|\alpha_v[i]|, |\alpha_v[i + 2^{(\lambda-1)}]|, |\alpha_v[i + 2^{\lambda}]|).
\label{eq:LLRl}
\end{aligned}
\end{equation}
When the hard decisions from the left branch ($\beta^t_{v_l}$) are computed, the LLRs can be proceed to the middle branch by
\begin{equation}
\begin{aligned}
\alpha^t_{v_c}[i] ={} & (1{\text -}2\beta^t_{v_l}[i])\alpha[i] + f^b(\alpha[i + 2^{(\lambda-1)}]+\alpha[i + 2^{\lambda}]).
\label{eq:LLRc}
\end{aligned}
\end{equation}
Finally, having $\beta^t_{v_l}$ and hard decisions from the middle branch ($\beta^t_{v_c}$), the LLR vector can travel to the right branch using
\begin{equation}
\begin{aligned}
\alpha^t_{v_r}[i] ={} & (1{\text -}2\beta^t_{v_l}[i])\alpha[i + 2^{(\lambda-1)}] + (1{\text -}2\beta^t_{v_l}[i]\oplus\beta^t_{v_c}[i])\alpha[i + \\2^{\lambda}].
\label{eq:LLRr}
\end{aligned}
\end{equation}
The hard decisions at node $\nu$ can be combined as
\begin{equation}
\begin{aligned}
\relax[\beta^{\nu t}_i, \beta^{\nu t}_{i + 2^{(\lambda-1)}},\beta^{\nu t}_{i + 2^{\lambda}}] ={} & [\beta^{\nu t_l}_{i}\oplus\beta^{\nu t_c}_{i}, \beta^{\nu t_l}_{i}\oplus\beta^{\nu t_r}_{i}, \beta^{\nu t_l}_{i}\oplus\\\beta^{\nu t_c}_{i}\oplus\beta^{\nu t_r}_{i}].
\label{eq:betak3}
\end{aligned}
\end{equation}
As illustrated in Fig. \ref{fig:EncDec} (b), we define (\ref{eq:LLRl}), (\ref{eq:LLRc}), (\ref{eq:LLRr}) and (\ref{eq:betak3}) as $f^t$, $g_1^t$, $g_2^t$ and $C^t$, respectively. Finally, a binary sign function ($s(x)$) and a frozen bit indicator vector ($a$) will be used in the following sections which are defined as 
\begin{equation}
\begin{aligned}
s(x) =     \begin{cases}
0,   & \text{if}\ l \geq 0, \\
1,   & \text{otherwise}
\end{cases},\
a_i =     \begin{cases}
0,   & \text{if}\ i \in \mathcal{F}, \\
1,   & \text{if}\ i \in \mathcal{I}.
\end{cases}
\label{sx}
\end{aligned}
\end{equation}
\section{MK codes: Construction and Architecture}
\label{sec_arch}
\subsection{Code Construction}
\begin{figure}
    \centering
    \includegraphics[width=1\columnwidth]{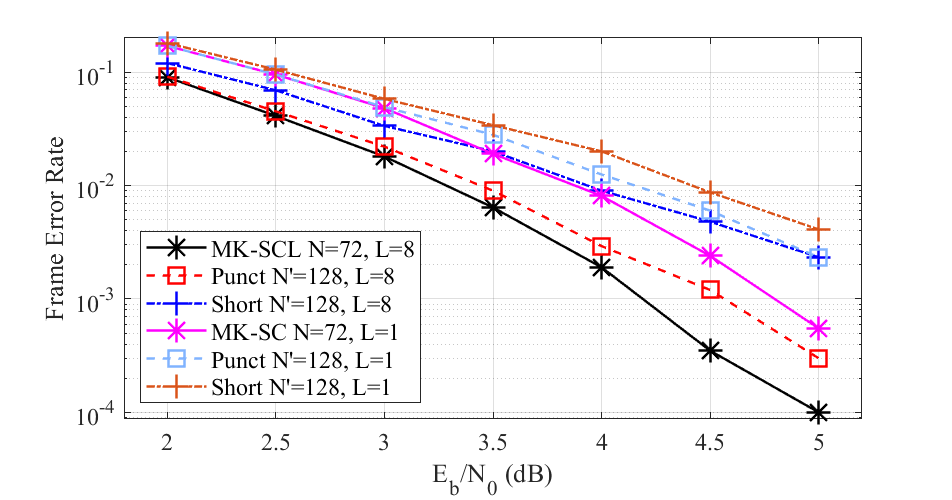}
    \caption{The error-correction performance of MK code of $\mathcal{PC}(72,36)$ comparing to puncturing and shortening methods.}
    \label{fig:MKCompSCL}
\end{figure}

\begin{figure}
    \centering
    \includegraphics[width=1\columnwidth]{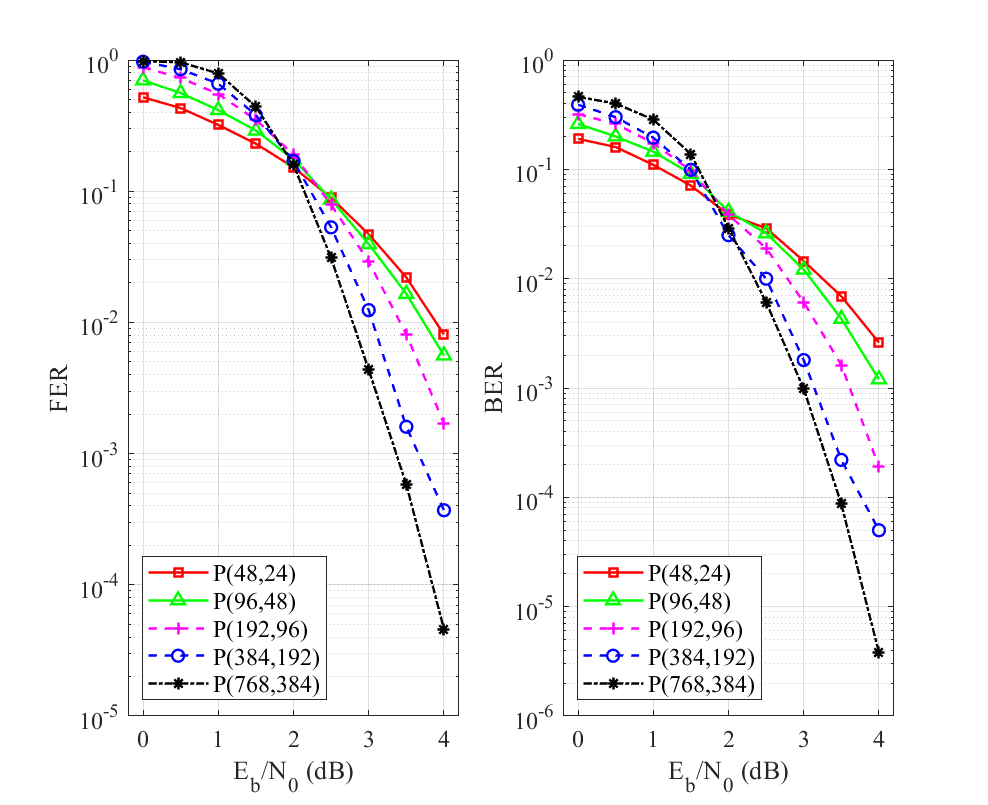}
    \caption{The error-correction performance of MK polar codes of rate $R = \frac{1}{2}$.}
    \label{fig:MKECC}
\end{figure}

In this paper, we use the method proposed in \cite{gabry2017} since it yields substantial error-correction performance comparing to puncturing \cite{Niu} and shortening \cite{Wang} methods. Fig. \ref{fig:MKCompSCL} plots  the error-correction performance of MK polar code of $\mathcal{PC}(72,36)$ with $G = T_3 \otimes T_2 \otimes T_2 \otimes T_2 \otimes T_3$. Obviously, it considerably outperforms punctured and shortened codes constructed by a mother code of $N^\prime = 128$. The MK codes can be constructed by arbitrary kernel orders. However, they feature different error-correction performances since the Kronecker product is not commutative. Presently, no theoretical way is identified to find the best kernel order and simulations need to be conducted to find the kernel order with the best error-correction performance. In this paper, we use the method proposed in \cite{bioglio} to obtain the kernel orders. We use the LDPC WiMAX code lengths \cite{Shin2012} as our guideline which suggests that the desired MK block lengths can be obtained by employing a few non-binary kernels. Thus we focus on MK codes constructed by only one ternary kernel. This is also important in terms of hardware complexity analysis as will be explained in the following sections. The error-correction performance of such codes with block lengths lower than $1024$ is illustrated in Fig. \ref{fig:MKECC}.

\begin{figure*}
    \centering
    \includegraphics[width=2\columnwidth]{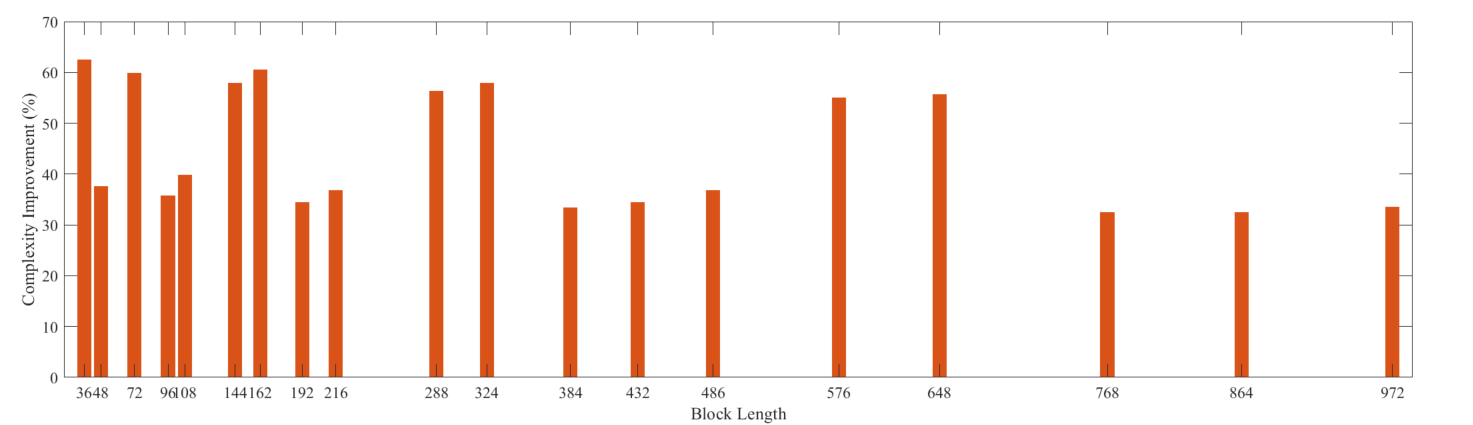}
    \caption{The complexity gain of MK method versus puncturing and shortening method using the defined complexity metric.}
    \label{fig:CompComparison}
\end{figure*}

The complexity of MK decoding is lower than that of the puncturing and shortening methods. This is due to the fact that MK polar codes employ a smaller Tanner graph with respect to  puncturing and shortening methods which use a mother code of size $N^\prime =2^{\lceil log_2N \rceil}$ which determines the code's complexity. The overall number of LLR computations can be defined as a complexity metric for the sake of comparison. Let $s$ be the number of stages in the code's Tanner graph which is identical to the number of kernels used in the code construction. The complexity metric of MK and puncturing/shortening methods can be calculated as $N\times s$ and $N^\prime log_2N^\prime$, respectively. 
Fig. \ref{fig:CompComparison} illustrates the complexity reduction of MK method with respect to puncturing and shortening methods for various code lengths. It can be seen that reported MK codes offer at least $32.5\%$ lower LLR computational complexity with respect to punctured/shortened codes.

\subsection{Proposed Decoder Architecture}
The SC decoder can be implemented by purely combinational logic since no loops are included in the algorithm. Therefore, there are no memory elements between the input and output stages. The main objective of combinational implementation is to achieve a high throughput. In this section, we first explain the implementation method of belief propagation functions. Then the overall architecture of the proposed mixed-kernel polar codes will be detailed. 
\subsubsection{Belief Propagation functions} To implement the combinational functions, $Q$ bits are used to represent the channel observation LLRs in sign-magnitude representation (similar to \cite{Dizdar} and \cite{Leroux}) to prevent conversions between different representations. The $f^b$ and $f^t$ functions can be directly implemented by (\ref{eq:sc_l}) and (\ref{eq:LLRl}), respectively, using comparators and multiplexers. The precomputation method in \cite{Zhang} is used to implement the $g^b$, $g_1^t$ and $g_2^t$ functions. 

The decision logic serves as the basic building block of the decoder. Using the construction method in \cite{bioglio}, we observed that all codes of this paper's interest has no ternary kernel located at stage zero ($s=0$ in Fig. \ref{fig:EncDec} (b)). Therefore, there is no need to use a ternary kernel in the decision logic considering the fact that they consume considerable resources. Thus we can use the method proposed in \cite{Dizdar} to estimate the binary odd-indexed leaves as 
\begin{equation}
\begin{aligned}
\hat{u}_{2i+1}=     \begin{cases}
0   & \text{if }\ a_{2i+1} = 0, \\
s(\lambda_1)   & \text{if } a_{2i+1} = 1\  \text{and}\ |\lambda_1| \geq |\lambda_0|\\
s(\lambda_0) \oplus  \hat{u}_{2i}  & \text{otherwise,}
\end{cases}
\label{uodd}
\end{aligned}
\end{equation}
where $\lambda_0$ and $\lambda_1$ are the input LLRs to the $g^b$ function.  

\subsubsection{Overall architecture} 
\begin{figure}
    \centering
    \includegraphics[width=1\columnwidth]{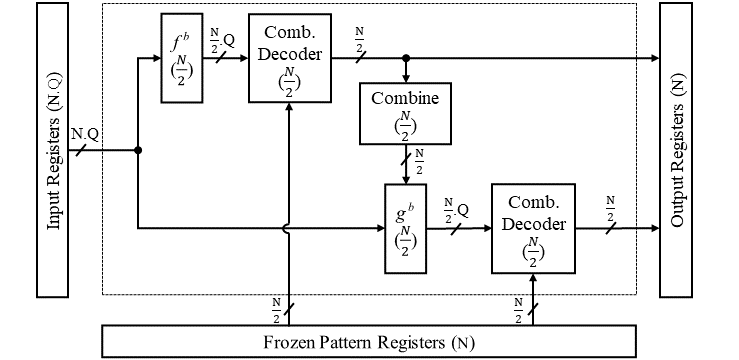}
    \caption{The combinational architecture of the Arikan's decoder.}
    \label{fig:K2Datapath}
\end{figure}

In this section, we first present an improved architecture for the Arikan's combinational decdoer proposed in \cite{Dizdar}. Then an SC decoding architecture for mixed-kernel polar codes will be proposed.
Fig. \ref{fig:K2Datapath} illustrates the generalized improved combinational architecture of a decoder of size $N$ constructed by Arikan's kernels based on the conventional architecture proposed in \cite{Dizdar}. In the conventional architecture, a decoder of size $N=4$ is used as the basic building block. However, we use a decoder of size $N=2$ to increase the code length flexibility. The modified decoder is constructed by two Arikan's decoders of size $N/2$ glued by one $f^b$, one $g^b$, and one combine logic of size $N/2$. The combine logic of size $N/2$ is used to substitute the encoder of size $N/2$ in the conventional architecture. This modification leads to significantly decreasing the number of XOR gates. For instance, for a polar code of size $N=32$, the number of XOR gates is decreased by $42,3\%$.

\begin{figure}
    \centering
    \includegraphics[width=1\columnwidth]{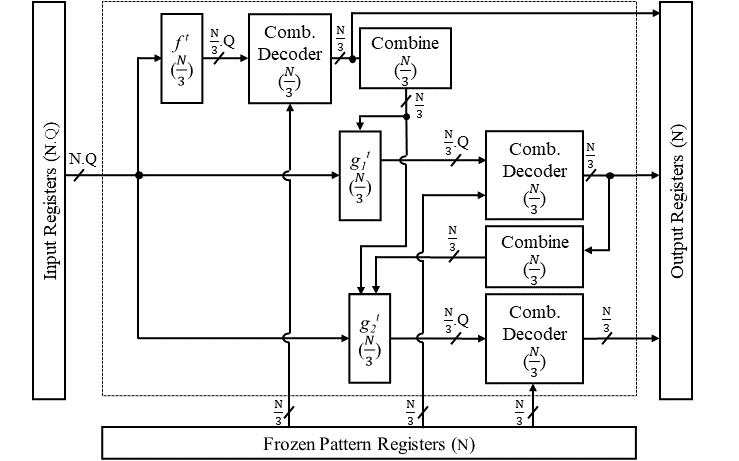}
    \caption{The combinational architecture of the pure-ternary decoder.}
    \label{fig:K3Datapath}
\end{figure}

The proposed combinational architecture of a ternary stage of a MK polar code of size $N$ is depicted in Fig. \ref{fig:K3Datapath}. It is composed of three decoders of size $N/3$ and one $f^t$, one $g_1^t$, one $g_2^t$ and two combine logics of size $N/3$ as the glue logic. The basic building block in this case is also a decoder of size $N=2$.

\begin{figure}
    \centering
    \includegraphics[width=1\columnwidth]{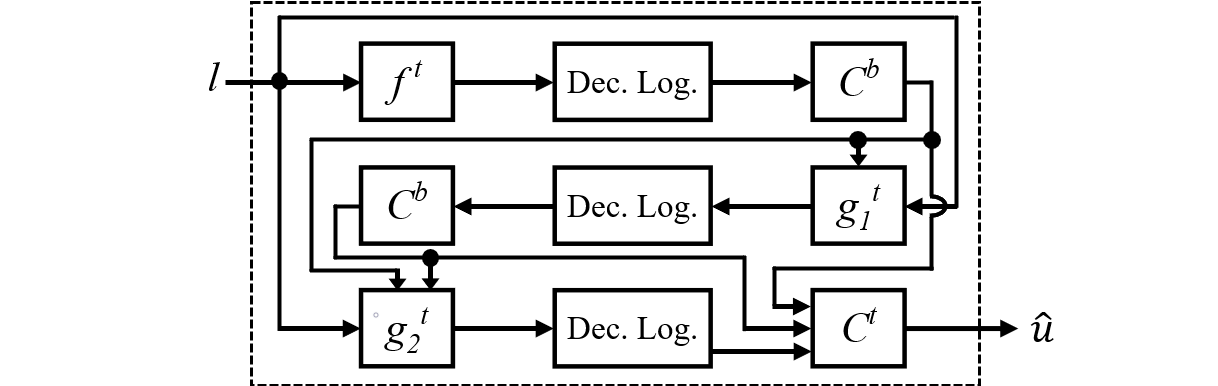}
    \caption{The combinational MK polar decoder for $N=6$ and $G = T_3 \otimes T_2$.}
    \label{fig:N6Datapath}
\end{figure}

As a result of recursive structure of the SC algorithm, the architecture of a MK decoder of size $N=6$ can be constructed by two combinational decoders of size $N=2$ and $N=3$. The last kernel which is used for polarizing the nodes located at stage zero ($s=0$ in Fig. \ref{fig:EncDec} (b)) known as decision logic serves as the basic building block of the MK decoder. Fig. \ref{fig:N6Datapath} portrays a mixed-kernel architecture for a polar code of size $N = 6$ with $G=T_3\otimes T_2$. Registers are not shown here to prevent congestion. Since the last kernel in the kernel sequence is $T_2$, a binary decision making circuitry is used as the basic building block of this decoder. Given $T_3$ as the next kernel, the glue logic includes two binary combine logics ($C^b$), one $f^b$, one $g_1^t$ and one $g_2^t$. In order to get the codeword estimate at the root of the tree, we use a ternary combine function ($C^t$ here) before writing the estimated codewords into the output registers. The first kernel determines whether to use a binary or ternary combine logic at the output stage of the decoder.

The Arikan's and proposed combinational architectures consume $N \times (Q+2)$ register bits to store the input LLRs ($N \times Q$), estimated codeword (N) and frozen pattern bits (N). As it is shown in Fig. \ref{fig:K2Datapath}, Fig. \ref{fig:K3Datapath} and Fig. \ref{fig:N6Datapath}, no synchronous logic elements (registers or RAM) are integrated between the input and output registers in the combinational decoders. This feature results in power efficiency and saving processing time. The hardware complexity is also decreased by removing the RAM routers and in result lowering the long read/write latencies. The decoder has one clock cycle latency since it generates the decision vector one clock after receiving the input LLRs. The critical path is equal to the delay of the logic between the input and output registers. 
\subsection{Hardware Complexity Analysis}
In this section, the complexity of the proposed architecture will be investigated. 
It is shown in \cite{Dizdar} that the total number of basic building blocks of a combinational Arikan's decoder of size $N$ can be computed as 
\begin{equation}
c_N^b+s_N^b+r_N^b = N(\frac{3}{2}log_2(N)-1)\approx Nlog_2(N),    
\label{complex_bin}
\end{equation}
where $c_N^b$ and $s_N^b$ denote the number of comparators used in implementing $f^b$ and the decision logic, respectively, and $r_N^b$ represents the total number of adders and subtractors used in implementing $g^b$.

A general complexity analysis of MK codes is not possible since it depends on the number and location of different kernels in the kernel sequence. However, in this paper we calculate the complexity of MK codes with only one ternary kernel since they construct the most important block lengths. We assume the most complex scenario where the ternary kernel is located at the root of the Tanner graph as the construction method in \cite{bioglio} suggests for the majority of such codes. The number of comparators ($c_N^{MK}$) used for implementing $f^t$ and $g_1^t$ equals $c_N^{MK}=N$. Therefore, a decoder of size $N$ has the recursive relationship of 
\begin{equation}
c_N^{MK} = 3c_{\frac{N}{3}}^{MK}+N=3(2c_{\frac{N}{6}}^b+\frac{N}{6})+N=....
\label{com_MK}
\end{equation}
Solving the recursion equation of (\ref{com_MK}) gives
\begin{equation}
c_N^{MK} = N(log_2N-4.17).
\label{com_1ter}
\end{equation}
The number of comparators used in the decision logic for $N=2$ is $s_2^b = 1$. Thus we can calculate the number of comparators in decision logic for MK case as $s_N^{MK} = \frac{N}{2}$. Finally, the total number of adders and subtractors ($r_N^{MK}$) can be computed as 
\begin{equation}
r_N^{MK} = 6c_{\frac{N}{3}}^b+\frac{4}{3}N = N(log_2N-1.25).
\label{com_MK2}
\end{equation}
Thus the number of basic logic blocks of the MK decoder can be estimated as 
\begin{equation}
c_N^{MK}+s_N^{MK}+r_N^{MK} = N(2 log_2N-4.92)\approx Nlog_2N.
\label{complex_ter}
\end{equation}
Which shows that the decoder's complexity is roughly $Nlog_2N$.
\section{Implementation Results and Analysis}
\label{sec_res}
\begin{figure}
    \centering
    \includegraphics[width=1\columnwidth]{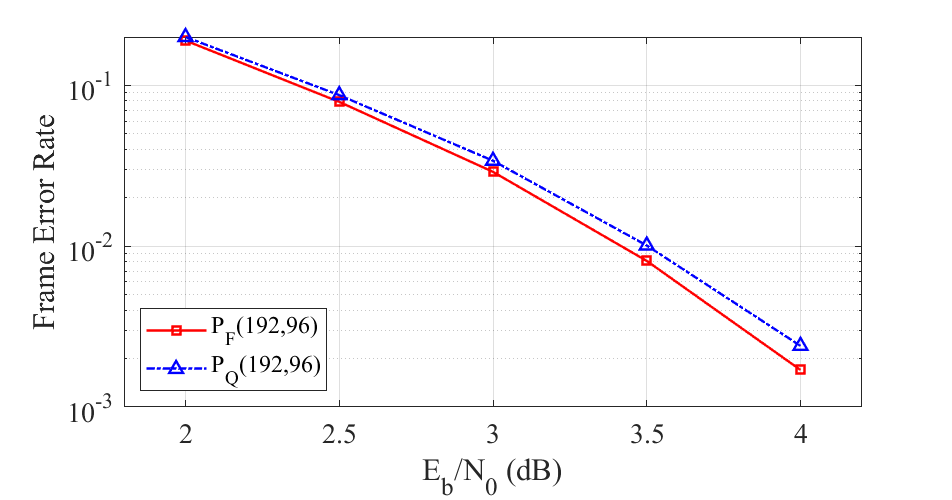}
    \caption{The error-correction performance of MK code of $\mathcal{PC}(192,96)$ comparing to puncturing and shortening methods.}
    \label{fig:MKQL}
\end{figure}

All polar codes of this paper are implemented by VHDL coding in Xilinx Vivado 2019.1 environment. Logic synthesis, technology mapping, and place and route are preformed to validate the design. Random codewords are generated by a software program and then transferred to the decoder. We use $Q=5$ bits to quantize the LLRs. Fig. \ref{fig:MKQL} illustrates that the quantization performance loss is negligible for $\mathcal{PC}(192,96)$.
\subsection{FPGA Utilization}
The FPGA utilization of the proposed MK decoders of size $N=48$ and $N=64$ along with that of a decoder of size $N=64$ proposed in \cite{Dizdar} are reported in Table \ref{tab:utilization}. Comparing to \cite{Dizdar}, the proposed decoder consumes $29,3\%$ lower LUTs. This is due to the modifications on the encoding circuitry in sub-decoders which results in using substantially lower XOR gates. The registers are mainly used for small logic circuits and fetching the RAM outputs. The proposed decoder consumes $5,8\%$ higher registers. This difference stems from register duplication to address the target clock frequency. Finally, both decoders consume the same amount of RAM since $N\times (Q+2)$ bits of RAM are needed.
\begin{table}
\centering
\caption{Post-fitting results of various polar codes.}
\begin{tabular}{cccccc}
\hline
Decoder&\begin{tabular}[c]{@{}c@{}}Block \\Length  \end{tabular}  &Kernel Order& LUTs  & Registers & \begin{tabular}[c]{@{}c@{}}RAM \\(bits)  \end{tabular}\\\hline
This work&48&\{3,2,2,2,2\}            &  2385 & 387  & 336    \\ 
This work&64&\{2,2,2,2,2,2\}          &  3626 & 415  & 448    \\
\cite{Dizdar}&64&\{2,2,2,2,2,2\}   &  5126 & 392  & 448 \\ 
\hline
\end{tabular}
\label{tab:utilization}
\end{table}
\subsection{Throughput}
The coded throughput of MK polar codes can be calculated as $\mathcal{T_C}=N.f$ similar to the binary case. 
The flexibility and scalability of the proposed decoder is evaluated by implementing different codes with different kernel orders. The coded throughput of various polar codes under different conditions are summarized in Table \ref{tab:throughput}. It can be observed that the proposed MK polar code of size $N=48$ achieves $22,4\%$ higher coded throughput with respect to combinational shortend code of the same size constructed by a mother code of $N^\prime=64$. Comparing to \cite{Coppolino}, our proposed MK decoder increases the throughput by $88,5\%$ for a polar code of length $N=48$.
The coded throughput of the proposed MK decoder of size $N=64$ gains $4\%$ higher throughput with respect to its Arikan's counterpart in \cite{Dizdar}.

\begin{table}
\centering
\caption{Coded throughput comparison of various polar codes}
\begin{tabular}{ccccccc}
\hline
\begin{tabular}[c]{@{}c@{}}Block \\Length  \end{tabular} &\begin{tabular}[c]{@{}c@{}}48 \\(MK)  \end{tabular}&\begin{tabular}[c]{@{}c@{}}48 \\(MK short.)  \end{tabular}&\begin{tabular}[c]{@{}c@{}}48\\(\cite{Coppolino} )  \end{tabular}  &\begin{tabular}[c]{@{}c@{}}64 \\(MK)  \end{tabular}&\begin{tabular}[c]{@{}c@{}}64 \\(Arikan \cite{Dizdar})  \end{tabular} \\\hline
$\mathcal{T_C}$ (Mbps) &812,1&663,45&430,9&884,6&850  \\
 \hline
\end{tabular}
\label{tab:throughput}
\end{table}
\section{Conclusion}
\label{sec_conc}
A combinational MK polar decoder with high-throughput has been implemented on FPGA. The proposed architecture offers flexible code rate. A complexity analysis is conducted and FPGA utilization for the target block lengths is reported. The implementation results reveal that the proposed architecture achieves the coded throughput of $812,1$ Mbps for a code of size $N=48$. 
\section*{Acknowledgment}
This research has been supported by the Academy of Finland, 6G Flagship program under Grant 346208.



\begin{thebibliography}{10}
\providecommand{\url}[1]{#1}
\csname url@samestyle\endcsname
\providecommand{\newblock}{\relax}
\providecommand{\bibinfo}[2]{#2}
\providecommand{\BIBentrySTDinterwordspacing}{\spaceskip=0pt\relax}
\providecommand{\BIBentryALTinterwordstretchfactor}{4}
\providecommand{\BIBentryALTinterwordspacing}{\spaceskip=\fontdimen2\font plus
\BIBentryALTinterwordstretchfactor\fontdimen3\font minus
  \fontdimen4\font\relax}
\providecommand{\BIBforeignlanguage}[2]{{%
\expandafter\ifx\csname l@#1\endcsname\relax
\typeout{** WARNING: IEEEtran.bst: No hyphenation pattern has been}%
\typeout{** loaded for the language `#1'. Using the pattern for}%
\typeout{** the default language instead.}%
\else
\language=\csname l@#1\endcsname
\fi
#2}}
\providecommand{\BIBdecl}{\relax}
\BIBdecl

\bibitem{Arikan}
E.~Arikan, ``Channel polarization: A method for constructing capacity-achieving
  codes,'' in \emph{2008 IEEE International Symposium on Information Theory},
  2008, pp. 1173--1177.

\bibitem{Rezaei2022}
H.~Rezaei, V.~Ranasinghe, N.~Rajatheva, M.~Latva-aho, G.~Park, and O.-S. Park,
  ``Implementation of ultra-fast polar decoders,'' in \emph{2022 IEEE
  International Conference on Communications Workshops (ICC Workshops)}, 2022,
  pp. 235--241.
  
\bibitem{9621127}
G.~Park, O.~-S.~Park, G.~Jo, H.~Rezaei, V.~Ranasinghe and N.~Rajatheva,
  ``Nonbinary polar codes constructions based on k-means clustering,'' in
  \emph{2021 International Conference on Information and Communication
  Technology Convergence (ICTC)}, 2021, pp. 640--643.

\bibitem{3GPP}
3rd Generation Partnership Project~(3GPP), \emph{5G; NR; Multiplexing and
  Channel Coding}.\hskip 1em plus 0.5em minus 0.4em\relax 3GPP document 38.212
  V.15.3.0, 2018.

\bibitem{Han2022}
S.~Han, B.~Kim, and J.~Ha, ``Rate-compatible punctured polar codes,''
  \emph{IEEE Communications Letters}, vol.~26, no.~4, pp. 753--757, 2022.

\bibitem{bioglio2017}
V.~Bioglio, F.~Gabry, and I.~Land, ``Low-complexity puncturing and shortening
  of polar codes,'' in \emph{2017 IEEE Wireless Communications and Networking
  Conference Workshops (WCNCW)}.\hskip 1em plus 0.5em minus 0.4em\relax IEEE,
  2017, pp. 1--6.

\bibitem{Xia2020}
C.~Xia, C.-Y. Tsui, and Y.~Fan, ``Construction of multi-kernel polar codes with
  kernel substitution,'' \emph{IEEE Wireless Communications Letters}, vol.~9,
  no.~11, pp. 1879--1883, 2020.

\bibitem{gabry2017}
F.~Gabry, V.~Bioglio, I.~Land, and J.-C. Belfiore, ``Multi-kernel construction
  of polar codes,'' in \emph{2017 IEEE International Conference on
  Communications Workshops (ICC Workshops)}.\hskip 1em plus 0.5em minus
  0.4em\relax IEEE, 2017, pp. 761--765.

\bibitem{bioglio}
V.~Bioglio, F.~Gabry, I.~Land, and J.-C. Belfiore, ``Minimum-distance based
  construction of multi-kernel polar codes,'' in \emph{GLOBECOM 2017-2017 IEEE
  Global Communications Conference}.\hskip 1em plus 0.5em minus 0.4em\relax
  IEEE, 2017, pp. 1--6.

\bibitem{Coppolino}
G.~Coppolino, C.~Condo, G.~Masera, and W.~J. Gross, ``A multi-kernel multi-code
  polar decoder architecture,'' \emph{IEEE Transactions on Circuits and Systems
  I: Regular Papers}, vol.~65, no.~12, pp. 4413--4422, 2018.

\bibitem{Sarkis}
G.~Sarkis, P.~Giard, A.~Vardy, C.~Thibeault, and W.~J. Gross, ``Fast polar
  decoders: Algorithm and implementation,'' \emph{IEEE Journal on Selected
  Areas in Communications}, vol.~32, no.~5, pp. 946--957, 2014.

\bibitem{Rezaei2022MK}
H.~Rezaei, N.~Rajatheva, and M.~Latva-Aho, ``Low-latency multi-kernel polar
  decoders,'' \emph{IEEE Access}, vol.~10, pp. 119\,460--119\,474, 2022.

\bibitem{Shin2012}
K.-W. Shin and H.~ju~Kim, ``A multi-mode ldpc decoder for ieee 802.16e mobile
  wimax,'' \emph{Journal of Semiconductor Technology and Science}, vol.~12, pp.
  24--33, 2012.

\bibitem{benammar2017}
M.~Benammar, V.~Bioglio, F.~Gabry, and I.~Land, ``Multi-kernel polar codes:
  Proof of polarization and error exponents,'' in \emph{2017 IEEE Information
  Theory Workshop (ITW)}.\hskip 1em plus 0.5em minus 0.4em\relax IEEE, 2017,
  pp. 101--105.

\bibitem{Niu}
K.~Niu, K.~Chen, and J.-R. Lin, ``Beyond turbo codes: Rate-compatible punctured
  polar codes,'' in \emph{2013 IEEE International Conference on Communications
  (ICC)}, 2013, pp. 3423--3427.

\bibitem{Wang}
R.~Wang and R.~Liu, ``A novel puncturing scheme for polar codes,'' \emph{IEEE
  Communications Letters}, vol.~18, no.~12, pp. 2081--2084, 2014.

\bibitem{Dizdar}
O.~Dizdar and E.~Arıkan, ``A high-throughput energy-efficient implementation
  of successive cancellation decoder for polar codes using combinational
  logic,'' \emph{IEEE Transactions on Circuits and Systems I: Regular Papers},
  vol.~63, no.~3, pp. 436--447, 2016.

\bibitem{Leroux}
C.~Leroux, A.~J. Raymond, G.~Sarkis, and W.~J. Gross, ``A semi-parallel
  successive-cancellation decoder for polar codes,'' \emph{IEEE Transactions on
  Signal Processing}, vol.~61, no.~2, pp. 289--299, 2013.

\bibitem{Zhang}
C.~Zhang and K.~K. Parhi, ``Low-latency sequential and overlapped architectures
  for successive cancellation polar decoder,'' \emph{IEEE Transactions on
  Signal Processing}, vol.~61, no.~10, pp. 2429--2441, 2013.

\end{thebibliography}
\end{document}